\documentstyle[jkas38]{article}

\runningauthor{KIM ET AL.}
\runningtitle{OPEN CLUSTERS: CZERNIK 24 AND CZERNIK 27} 
\beginpage{1}
\endpage{7}

\newcommand{\hii}{{\sc H~ii}\/ }

\def\simlt{\lower.5ex\hbox{$\; \buildrel < \over \sim \;$}}
\def\simgt{\lower.5ex\hbox{$\; \buildrel > \over \sim \;$}}
\def\arcdeg{\hbox{$^\circ$}}
\def\arcmin{\hbox{$^\prime$}}
\def\arcsec{\hbox{$^{\prime\prime}$}}
\slugcomment{Not to appear in Nonlearned J., 45. \today}

\begin{document}

%\title{BOAO Photometric Survey of Galactic Open Clusters. III.
%  Czernik 24 and Czernik 27}
\title{BOAO PHOTOMETRIC SURVEY OF GALACTIC OPEN CLUSTERS. III.
   CZERNIK 24 AND CZERNIK 27}
\author{Sang Chul Kim$^1$, Hong Soo Park$^2$, Sangmo Tony Sohn$^1$, 
   Myung Gyoon Lee$^2$, Byeong-Gon Park$^1$, \\
   Hwankyung Sung$^3$, 
   Hong Bae Ann$^4$, Moo-Young Chun$^1$, Seung-Lee Kim$^1$, 
   Young-Beom Jeon$^1$, In-Soo Yuk$^1$, and Sang Hyun Lee$^5$}

\offprints{S. C. Kim}
\address{$^1$ Korea Astronomy and Space Science Institute, Daejeon 305-348, Korea; \\
  (sckim, tonysohn, bgpark, mychun, slkim, ybjeon, yukis@kasi.re.kr)}
\address{$^2$ Astronomy Program, SEES, Seoul National University, Seoul 151-742, Korea; \\
  (hspark@astro.snu.ac.kr, mglee@astrog.snu.ac.kr)}
\address{$^3$ Department of Astronomy and Space Science, Sejong University, 
   Seoul 143-747, Korea; (sungh@sejong.ac.kr)}
\address{$^4$ Department of Earth Science, Pusan National University, Pusan 609-735, Korea;
  (hbann@pusan.ac.kr)}
\address{$^5$ Gimhae Astronomical Observatory, Gimhae 621-170, Korea;
   (ngc2420@gsiseol.or.kr)}
%\address{\it E-mail: sckim@kasi.re.kr}

\vskip 3mm
\address{\normalsize{\it (Received Nov. 4, 2005; Accepted Nov. 18, 2005)}}

\abstract{
We present $BV$ CCD photometry 
  for the open clusters Czernik 24 and Czernik 27.
These clusters have never been studied before, 
  and we provide, for the first time, the cluster parameters; 
  reddening, distance, metallicity and age.
Czernik 24 is an old open cluster with age $1.8 \pm 0.2$ Gyr,
  metallicity [Fe/H]$=-0.41 \pm 0.15$ dex, 
  distance modulus ${\rm (m-M)}_0 = 13.1 \pm 0.3$ mag (d$=4.1 \pm 0.5$ kpc), and
  reddening $E(B-V) = 0.54 \pm 0.12$ mag.
The parameters for Czernik 27 are estimated to be age = $0.63 \pm 0.07$ Gyr,
  [Fe/H]$=-0.02 \pm 0.10$ dex,
  ${\rm (m-M)}_0 = 13.8 \pm 0.2$ mag (d$=5.8 \pm 0.5$ kpc), and
  $E(B-V) = 0.15 \pm 0.05$ mag.
The metallicity and distance values for Czernik 24 are consistent with 
  the relation between the metallicity and the Galactocentric distance 
  of other old open clusters. 
We find the metallicity gradient of 51 old open clusters including Czernik 24 
  to be $\Delta$[Fe/H]/$\Delta R_{gc}= -0.064 \pm 0.009$ dex kpc$^{-1}$.
}


\maketitle
\section{INTRODUCTION}

Open clusters (OCs) are an excellent probe for the study of 
  the structure and evolution of the Galactic disk (Friel 1995).
While Lyng\r{a} (1987) has published some parameters for over 1200 OCs %Lynga
  and Dias et al. (2002, http://www.astro.iag.usp.br/$\sim$wilton/,
  version 2.5, 2005 October 3) have presented data for 1753 OCs,
  only about a few hundreds of OCs have 
  photometry good enough for the estimation of the physical parameters of the clusters
  (Friel 1993; Phelps, Janes, \& Montgomery 1994).
Ann et al. (1999) and Ann et al. (2002) have presented the results from 
  the BOAO (Bohyunsan Optical Astronomy Observatory) photometric survey 
  of Galactic OCs which is aimed at observing clusters 
  with few, if any, previous studies,
  and this paper is the third in this series.

The OCs Czernik 24 and Czernik 27 were first identified by Czernik (1966),
  and there have been no photometric study for either of these two clusters.
The preliminary results presented in a conference proceedings 
  (Park et al. 2001) are superseded by this paper.
In this paper we present the first photometric study of 
  Czernik 24 and Czernik 27, and provide the basic parameters.
Section II describes the observations and data reduction.
Section III and IV present the analysis for Czernik 24 and Czernik 27,
  respectively.
Section V discusses the results and 
  a summary and conclusions are given in Section VI.

\section{OBSERVATIONS AND DATA REDUCTION}
\subsection{Observations}
$BV$ CCD images of Czernik 24 and Czernik 27 were obtained 
  using the SITe 2048 $\times$ 2048 CCD camera
  (24$\micron$ pixel)
  at the BOAO 1.8 m telescope on 2000 December 26.
The journal of observations of Czernik 24 and Czernik 27 is
  given in Table 1.
The field of view of the CCD image is $11.\arcmin7 \times 11.\arcmin7$
  and the pixel scale is 0.34\arcsec~pixel$^{-1}$
  at the f/8 Cassegrain focus of the telescope.
The gain and readout noise are, respectively, 1.8 electrons per ADU
  and 7 electrons.

Color maps of Czernik 24 and Czernik 27 are illustrated in
  Figure 1, 
  which shows that both objects are loose clusters.
The centers of Czernik 24 and Czernik 27 are estimated approximately
  to be at 
  R.A.(2000)$ = 05^h~ 55^m~ 27^s$ and
  Decl.(2000) $ = +20\arcdeg~ 52\arcmin~ 59\arcsec$
  ($X=444$ pixel and $Y=1107$ pixel) and
  R.A.(2000)$=07^h~ 03^m~ 22^s$ and
  Decl.(2000) $ =+06\arcdeg~ 23\arcmin~ 47\arcsec$
  ($X=617$ pixel and $Y=898$ pixel), respectively.
The approximate radii of Czernik 24 and Czernik 27 are estimated 
  to be 290 pixels ($\sim 99\arcsec$) and 
  520 pixels ($\sim 177\arcsec$), respectively.

%-------------------------------------------------- FIG 1 START
\begin{figure}[p]
  \epsfxsize=7.5cm
  \epsfysize=7.5cm
  \centerline{\epsffile{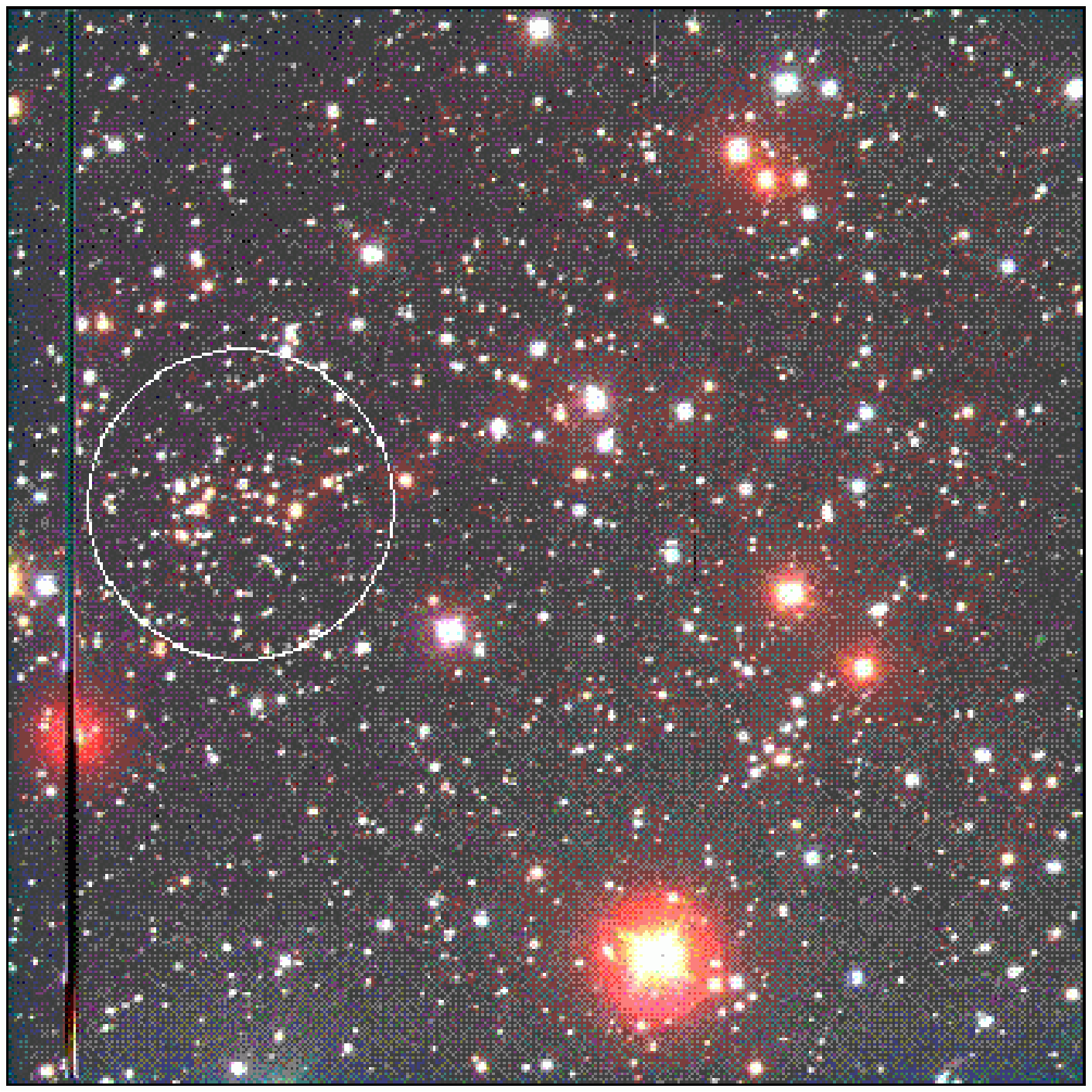}}
  \epsfxsize=7.5cm
  \epsfysize=7.5cm
  \centerline{\epsffile{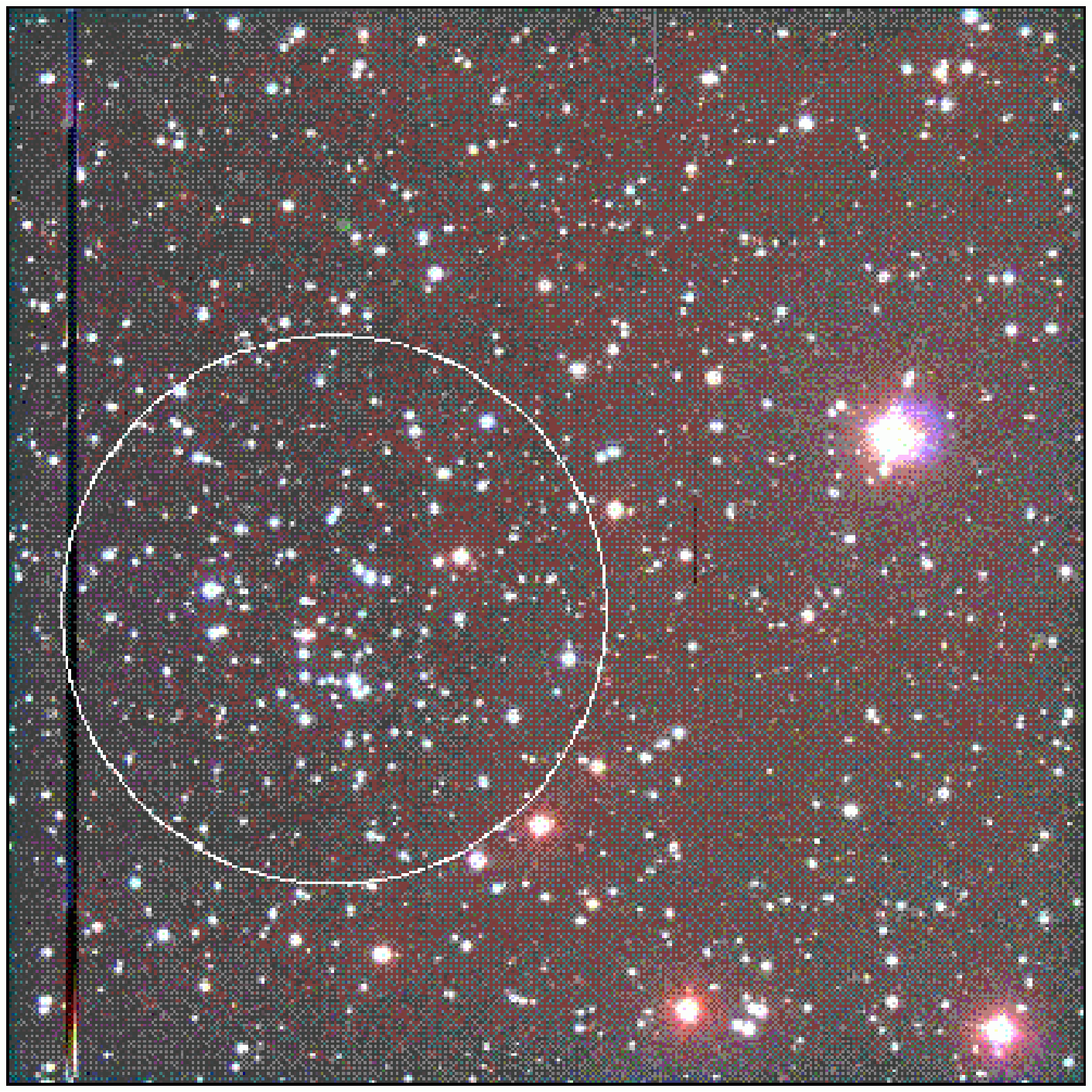}} 
{\small {\bf ~~~Fig. 1.}---~Color images of Czernik 24 (upper panel)
and Czernik 27 (lower panel). 
The approximate sizes of the clusters are denoted as circles
  with radius of $\sim 99\arcsec$ and $\sim 177\arcsec$ for
  Czernik 24 and Czernik 27, respectively.
North is at the top and east is to the left.
The size of each field is 11\arcmin.7 $\times$ 11\arcmin.7.
}
\end{figure}
%-------------------------------------------------- FIG 1 END

% -------------------------------------------------- TABLE 1 START
\begin{table}[t] 
\begin{center} 
{\bf Table 1.}~~Observation Log \\
\vskip 3mm
{\small
\setlength{\tabcolsep}{1.2mm} 
\begin{tabular}{cccc} \hline\hline
Cluster & Filter & ${\rm T_{exp}}$ & Seeing \\
\hline
Czernik 24 & $B$ & 300 s, 30 s & 1.\arcsec7 \\
           & $V$ & 150 s, 12 s & 1.\arcsec8 \\
Czernik 27 & $B$ & 300 s, 30 s & 2.\arcsec0 \\
           & $V$ & 150 s, 15 s & 1.\arcsec9 \\
\hline
\end{tabular}
} %\small
\end{center}
\end{table}
%-------------------------------------------------- TABLE 1 END

\subsection{Data Reduction}
The original CCD images were bias subtracted and flat-fielded 
  using the standard CCDPROC task within IRAF\footnote{IRAF is 
  distributed by the National
  Optical Astronomy Observatories, which are operated by the Association of
  Universities for Research in Astronomy, Inc., under cooperative agreement
  with the National Science Foundation.}.  
Instrumental magnitudes of the stars
  in the CCD images were obtained using the digital stellar photometry
  reduction program IRAF/{\small DAOPHOT} (Stetson 1987; Davis 1994).  
The resulting instrumental magnitudes were transformed on to 
  the standard system using the standard stars %SA 98 and Rubin 149 area 
  (Landolt 1992) observed on the same night.  
The night of observation was considered partly semi-photometric,
  so we secured our photometry using a secondary method as following.  
OC Berkeley 29 was observed on the same night 
  we obtained images of Czernik 24 and Czernik 27.  
We derived standard $BV$ magnitudes for stars in our  Berkeley 29 field 
  and compared with those of Kaluzny (1994)\footnote{We note that 
  the $BVI$ magnitudes of Kaluzny (1994) and Tosi et al. (2004) are 
  in excellent agreement (see Figure 3 of Tosi et al. 2004).} 
  as shown in Figure 2.
Using stars in the range $15.0 < B_{BOAO} < 19.0$ and 
  $14.5 < V_{BOAO} < 18.0$, we computed offsets of 
  $\Delta B = 0.276 \pm 0.021$ and $\Delta V = 0.212 \pm 0.013$ 
  (this study $-$ Kaluzny).  
Finally, these magnitude offsets were applied to our
  photometry of Czernik 24 and Czernik 27 fields.

%-------------------------------------------------- FIG 2 START
\begin{figure}

\centerline{\epsfxsize=8.1cm\epsfbox{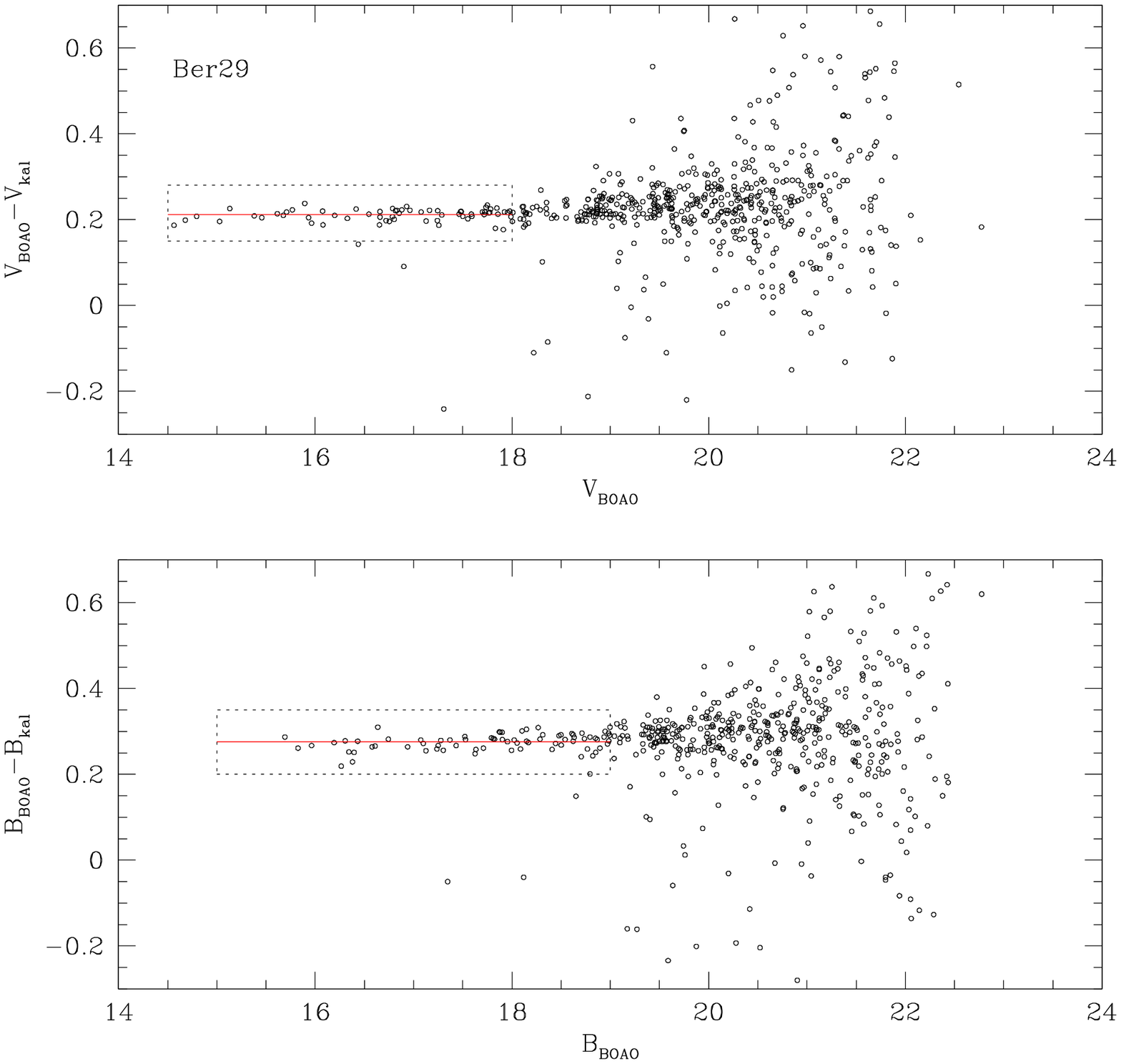}}
{\small {\bf ~~~Fig. 2.}---~Comparison of our $B, V$ photometry 
  with the one in Kaluzny (1994).
The differences of the two studies were estimated to be
  $\Delta B=0.276 \pm 0.021$ and $\Delta V=0.212 \pm 0.013$
  (this study $-$ Kaluzny) 
  from the box regions of dotted lines.
}
\end{figure}
%-------------------------------------------------- FIG 2 END

%=======================================================================
\section{ANALYSIS FOR CZERNIK 24}
\subsection{Color-Magnitude Diagrams}
Figure 3 shows the $V-(B-V)$ color-magnitude diagram (CMD)
  of the measured stars in the observed region in Czernik 24.
The CMD of the central region (lower panel) consists mostly of 
  the members of Czernik 24 
  with some possible contamination of field stars.
Some noticeable features seen in the CMDs
  of Czernik 24 are:
(i) a well-defined main sequence, the top of which is located at
  $V \approx 16.5$ mag;
(ii) a gap at $V \approx 17.7$ mag in the main sequence;
(iii) a few red giant clump (RGC) stars on the red giant branch sequence
  around $(B-V) \approx 1.4$ and $V \approx 15.5$ mag,
  which is noted by squares in Figure 3; and
(iv) few stars along the locus of the red giant branch.

%-------------------------------------------------- FIG 3 START
\begin{figure}
\centerline{\epsfxsize=7.1cm\epsfbox{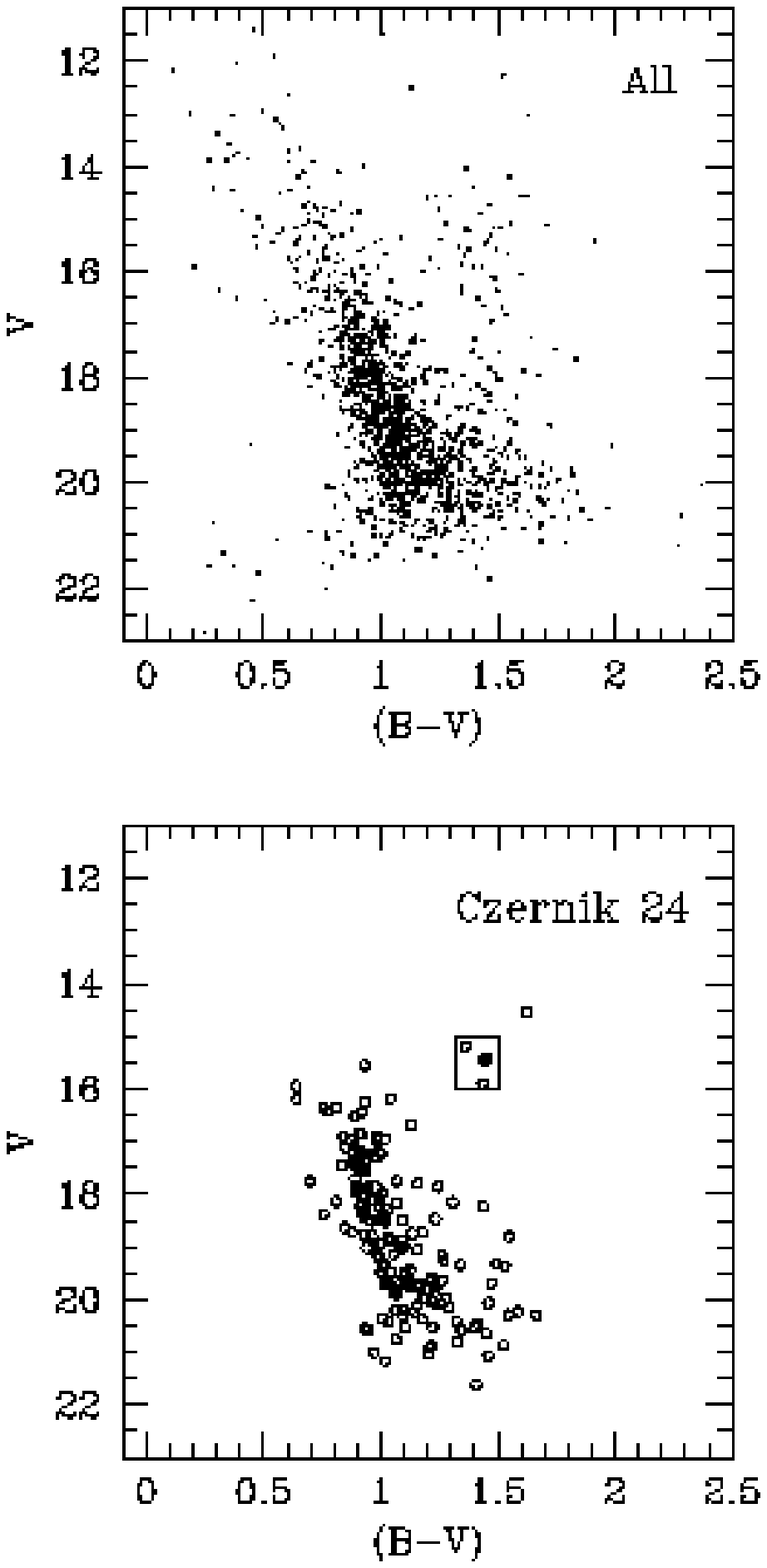}}
{\small {\bf ~~~Fig. 3.}---~$V-(B-V)$ color-magnitude diagrams
  of Czernik 24.
Upper panel is for the entire region, and 
  lower panel is for the central region of Czernik 24
  denoted as a circle in Figure 1 (r $< 100\arcsec$).
The square represents the position of the red giant clump.
}
\end{figure}
%-------------------------------------------------- FIG 3 END

\subsection{Reddening}
Since Czernik 24 is located at low Galactic latitude ($b = -2.\arcdeg21$),
  it is expected that the interstellar reddening toward Czernik 24
  could be significant. 
We have estimated the reddening toward Czernik 24 using the mean color
  of the RGC stars.
It is known that the magnitudes and colors of the RGC stars in old open clusters
   show small variations (Cannon 1970; Janes \& Phelps 1994). 
Janes \& Phelps (1994) estimated the mean color and magnitude
  of the RGC in old open clusters to be
  $(B-V)_{0, RGC} = 0.87 \pm 0.12$, and $M_{V, RGC} = 0.59 \pm 0.46$,
  when the $V$ magnitude difference between the RGC and
  the main-sequence turn-off of the clusters, $\delta V$, is less than one. %MSTO
The mean color and the mean magnitude of the RGC of Czernik 24 are estimated to be
  $(B-V)_{RGC} = 1.41 \pm 0.01$ and $V_{RGC} = 15.50 \pm 0.05$ mag, respectively,
  while the main-sequence turn-off magnitude and $\delta V$ are estimated to be 
  $16.4 \pm 0.1$ mag $0.9 \pm 0.2$ mag, respectively.
Therefore, the resulting reddening value is estimated to be 
  $E(B-V) = (B-V)_{RGC} - (B-V)_{0,RGC} = 0.54 \pm 0.12$.

\subsection{Isochrone Fitting}
We derive the cluster parameters by fitting the theoretical isochrones
  given by the Padova group (Girardi et al. 2000).
Figure 4 shows the best matched isochrone,
  which gives age$=1.8 \pm 0.2$ Gyr,
  [Fe/H] $=-0.41 \pm 0.15$ dex, and 
  the true distance modulus ${\rm (m-M)}_0 = (V-M_V) -3.1 \times E(B-V)
  = 13.1 \pm 0.3$ (d $= 4.1 \pm 0.5$ kpc).
The derived age is consistent with the presence of RGC,
  confirming that Czernik 24 is a cluster consisting of stars
  that evolved away from the main sequence.
  
%-------------------------------------------------- FIG 4 START
\begin{figure}

\centerline{\epsfxsize=7.1cm\epsfbox{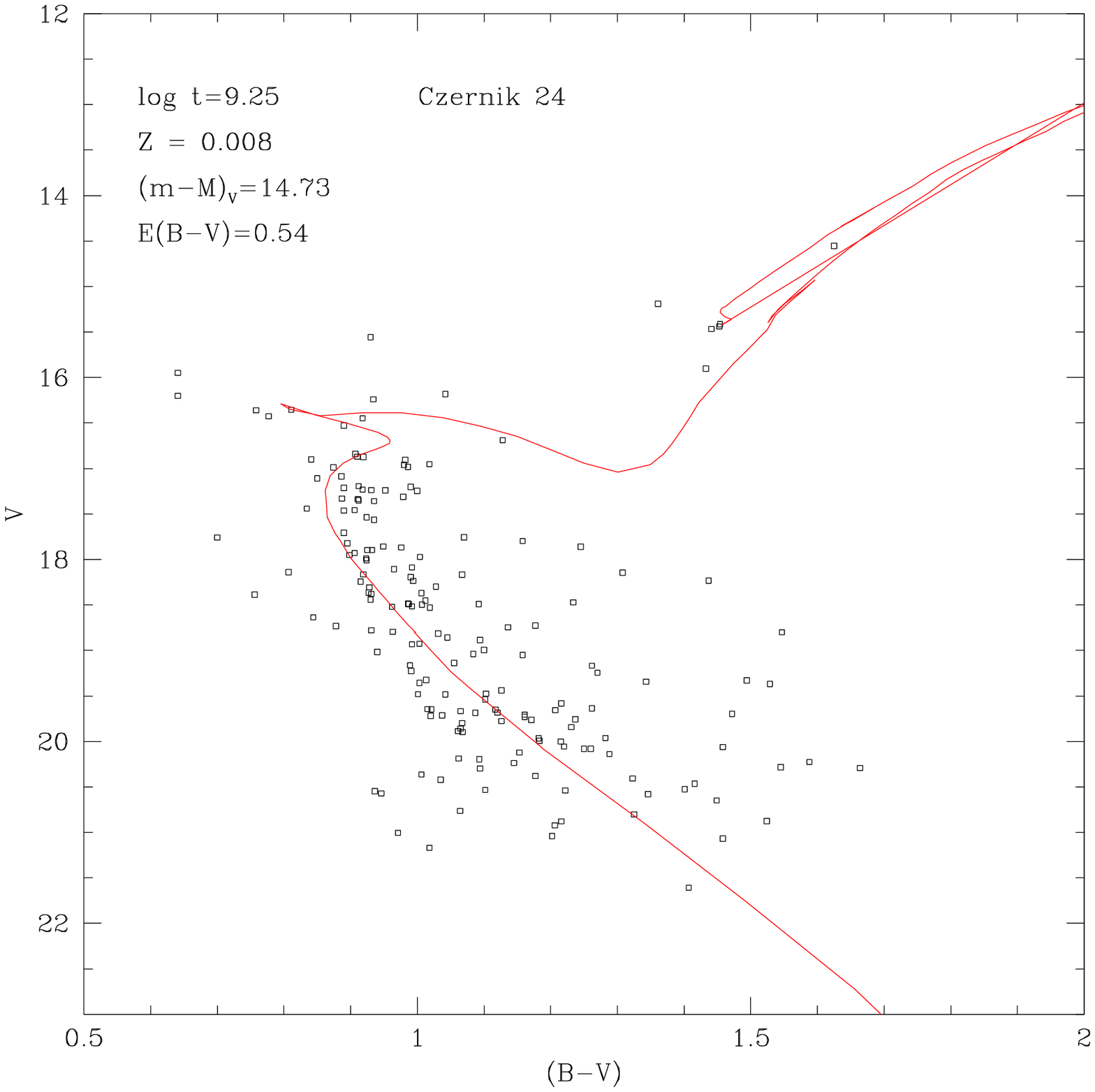}}
{\small {\bf ~~~Fig. 4.}---~Isochrone fitting for Czernik 24
  in the $V-(B-V)$ color-magnitude diagram.
The solid line represents the Padova isochrone for age = 1.8 Gyr,
  [Fe/H] $=-0.41$ dex, shifted according to the reddening and distance 
  of Czernik 24.
}
\end{figure}
%-------------------------------------------------- FIG 4 END

%=======================================================================
\section{ANALYSIS FOR CZERNIK 27}
\subsection{Color-Magnitude Diagrams}
Figure 5 shows the $V-(B-V)$ CMD
  of the measured stars in the observed region in Czernik 27.
The CMD of the central region (lower panel) consists mostly of 
  the members of Czernik 27 
  with some possible contamination of the field stars.
Some noticeable features seen in the CMDs
  of Czernik 27 are:
(i) a well-defined main sequence, the top of which is located at
  $V \approx 14.8$ mag;
(ii) a gap at $V \approx 16.2$ mag in the main sequence; and
(iii) only a few stars along the locus of the red giant branch.
The four stars above the top of the main sequence could be
  blue stragglers or foreground field stars,
  which needs further study to clarify.

%-------------------------------------------------- FIG 5 START
\begin{figure}
\centerline{\epsfxsize=7.1cm\epsfbox{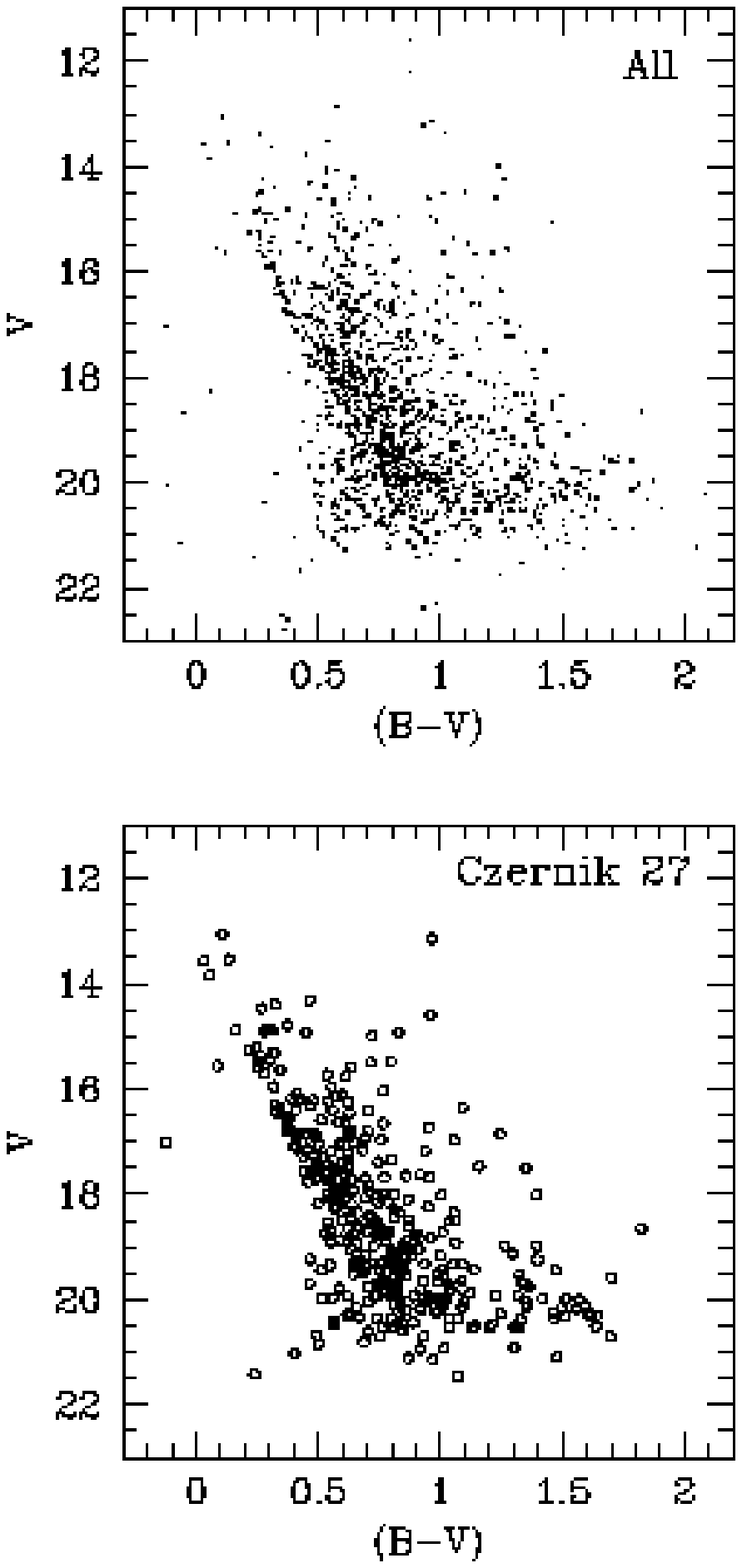}}
{\small {\bf ~~~Fig. 5.}---~$V-(B-V)$ color-magnitude diagrams
  of Czernik 27.
Upper panel is for the entire region, and 
  lower panel is for the central region of Czernik 27
  denoted as a circle in Figure 1 (r $\le 177\arcsec$).
}
\end{figure}
%-------------------------------------------------- FIG 5 END

\subsection{Isochrone Fitting}
Since Czernik 27 is also located at low Galactic latitude ($b = +5.\arcdeg56$),
  it is expected that the interstellar reddening toward Czernik 27
  could be significant.
Unlike in the case of Czernik 24 where there are RGC stars,
  it is not possible to use the same method 
  to estimate the reddening toward Czernik 27 
  since there are no RGC stars.

We derive the cluster parameters by fitting the theoretical isochrones
  given by the Padova group (Girardi et al. 2000).
Figure 6 shows the best matched isochrone,
  which gives age$=0.63 \pm 0.07$ Gyr,
  [Fe/H] $=-0.02 \pm 0.10$ dex, 
  the true distance modulus ${\rm (m-M)}_0 = (V-M_V) -3.1 \times E(B-V) 
  = 13.8 \pm 0.2$ (d $= 5.8 \pm 0.5$ kpc), and 
  the reddening $E(B-V)=0.15 \pm 0.05$.
The rather younger age of Czernik 27 than that of Czernik 24 is
  consistent with the fact that 
  Czernik 27 has no evolved stars like RGC stars and
  has only a few stars along the red giant branch.
If the four stars above the top of the main sequence are 
  members of the main sequence, then the age might be still younger.

%-------------------------------------------------- FIG 6 START
\begin{figure}
\centerline{\epsfxsize=7.1cm\epsfbox{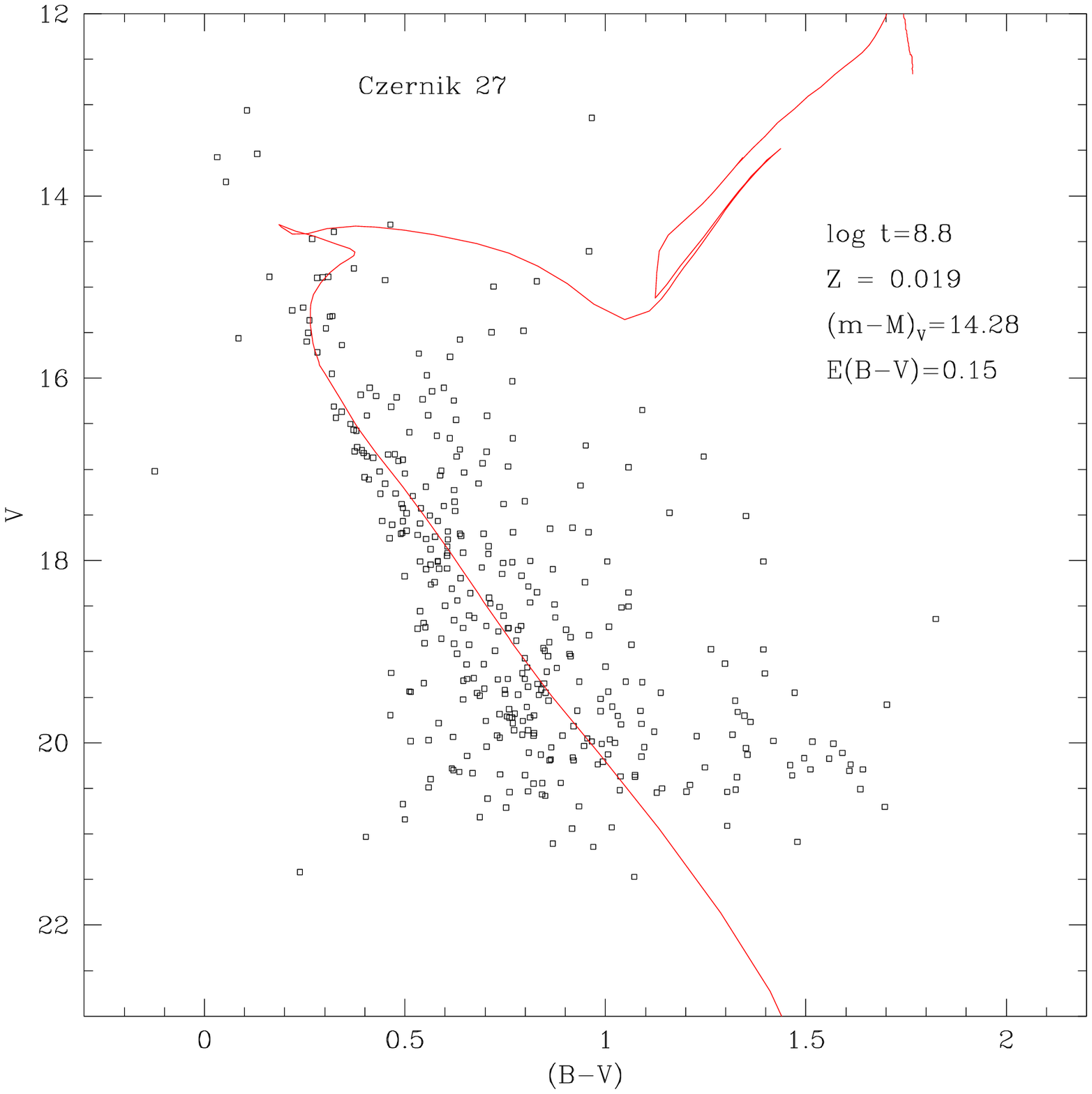}}
{\small {\bf ~~~Fig. 6.}---~Isochrone fitting for Czernik 27
  in the $V-(B-V)$ color-magnitude diagram.
The solid line represents the Padova isochrone for age = 0.63 Gyr,
  [Fe/H] $=-0.02$ dex, shifted according to the reddening and distance 
  of Czernik 27.
}
\end{figure}
%-------------------------------------------------- FIG 6 END

\section{DISCUSSION}
It is generally believed that the old open clusters in the Galactic disk
  show metallicity gradient of decreasing metallicities
  along the Galactocentric distances.
It is also supported by the studies of other metallicity tracers 
  such as \hii regions, bright blue stars, red giants,
  and planetary nebulae (Portinari \& Chiosi 1999;
  Hou, Prantzos, \& Boissier 2000).
Various estimates of the metallicity gradient 
  from previous studies on open clusters
  are summarized in Table 2.
From a sample of 69 young OB stars, 
  Daflon \& Cunha (2004) have obtained a similar metallicity gradient of
  $-0.042 \pm 0.007$ dex kpc$^{-1}$.

% -------------------------------------------------- TABLE 2 START
\begin{table}[t]
\begin{center}
{\bf Table 2.}~~Slope of the Galactocentric Radial [Fe/H] Gradient\\
\vskip 3mm
{\small
\setlength{\tabcolsep}{1.2mm}
\begin{tabular}{cc} \hline\hline
Paper & $\Delta$[Fe/H]/$\Delta R_{gc}$ \\ %& Comment \\
      & (in dex kpc$^{-1}$)     \\ %&         \\
\hline
Friel (1995)                       & $-0.091 \pm 0.014$ \\
Carraro, Ng, \& Portinari (1998)   & $-0.09$            \\
Park \& Lee (1999)                 & $-0.086 \pm 0.011$ \\
Friel (1999)                       & $-0.06  \pm 0.01 $ \\
Friel et al. (2002)                & $-0.059 \pm 0.010$ \\
Chen, Hou, \& Wang (2003)          & $-0.063 \pm 0.008$ \\
Kim \& Sung (2003)                 & $-0.064 \pm 0.010$ \\
Salaris, Weiss, \& Percival (2004) & $-0.055 \pm 0.019$ \\
Carraro et al. (2004)              & $-0.03  \pm 0.01 $ \\ 
\hline
Mean value                         & $-0.066 \pm 0.019$ \\
\hline
\end{tabular}
} %\footnotesize
\end{center}
\end{table}
%-------------------------------------------------- TABLE 2 END

Friel et al. (2002) have suggested a hint of slight steepening of 
  the abundance gradient with increasing cluster age, %their Fig 3
  while Salaris et al. (2004) have obtained the opposite trend.
As noted by Salaris et al. (2004), it is necessary 
  to increase the cluster sample size with [Fe/H] and distance 
  on a homogeneous scale
  for a better determination of the dependency of the radial metallicity gradient
  on age.
Twarog, Ashman, \& Anthony-Twarog (1997) have suggested a sharp discontinuity
  in the radial metallicity distribution at $R_{gc} = 10$ kpc, %their Fig 3
  which is still in debate (e.g., Bragaglia et al. 2000).

Using the data of 50 old open clusters compiled in Kim \& Sung (2003,
  references therein) and those of Czernik 24\footnote{Note that 
  Czernik 27 is not an old open cluster.}
  obtained in this study,
  we have plotted the open cluster metallicity [Fe/H] versus
  Galactocentric distance in Figure 7.
We adopted the Galactocentric distance of the Sun to be 8.5 kpc.
Czernik 24 is shown as a filled diamond in this plot,
  based on the parameters obtained in this study.
The solid line represents a least-squares fit to the data that yields an
  [Fe/H] gradient of $\Delta$[Fe/H]/$\Delta R_{gc}=-0.064 \pm 0.009$ dex kpc$^{-1}$.
This value is in good agreement with the values obtained in other studies
  (see Table 2),
  and we find that the metallicity and the Galactocentric distance of
  Czernik 24 are consistent with the general trend
  obtained from the studies on old open clusters.

%-------------------------------------------------- FIG 7 START
\begin{figure}
\centerline{\epsfxsize=7.1cm\epsfbox{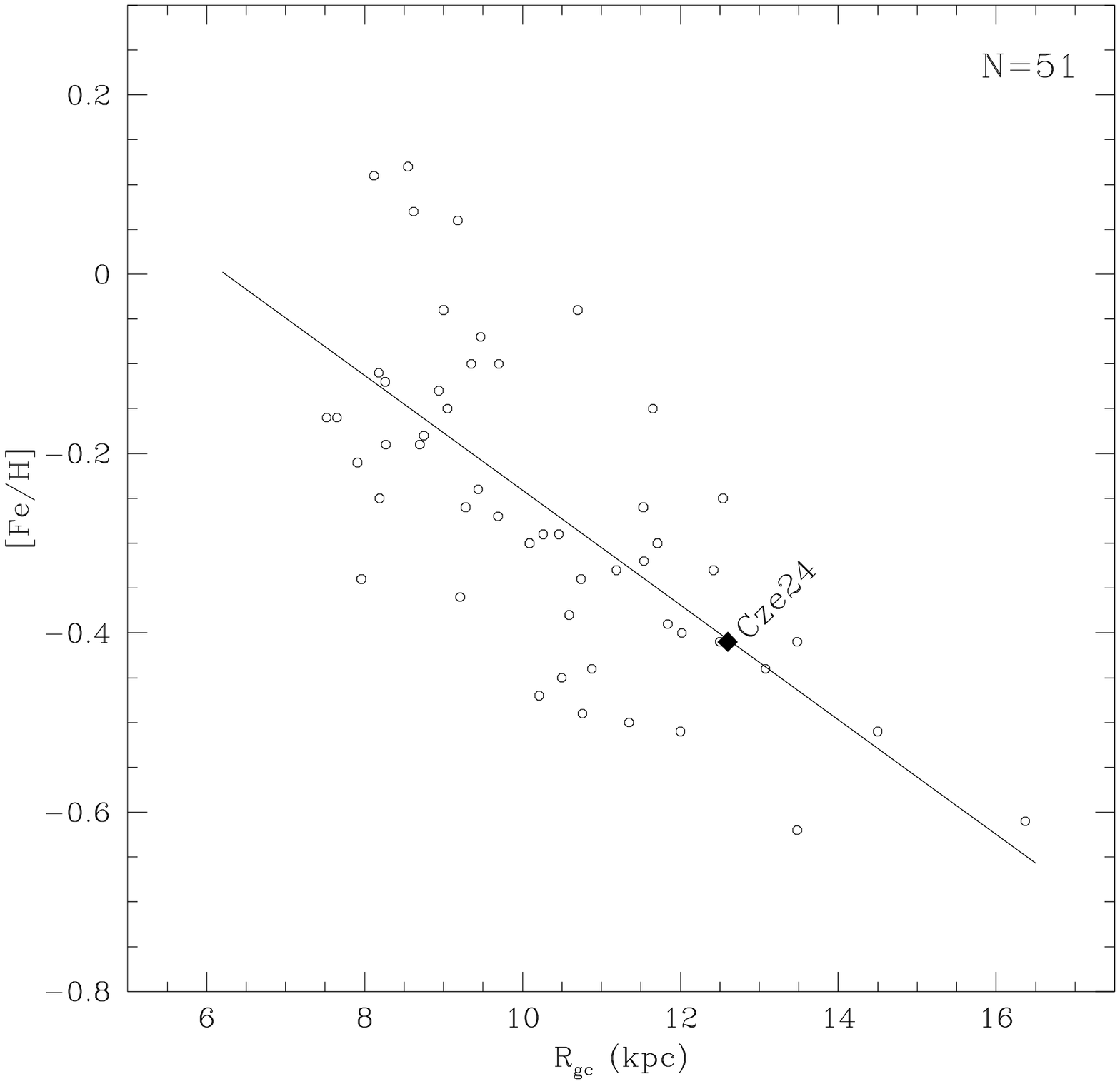}}
{\small {\bf ~~~Fig. 7.}---~Radial abundance gradient for 51 old open clusters.
The solid line is a least-squares fit to the data that yields an
abundance gradient of $\Delta$[Fe/H]/$\Delta R_{gc}=-0.064 \pm 0.009$ dex kpc$^{-1}$.
Filled square is the position of Czernik 24 based
  on the parameters obtained in this study.
}
\end{figure}
%-------------------------------------------------- FIG 7 END

\section{SUMMARY AND CONCLUSIONS}
We have presented the analysis of the photometry of 
  the old open clusters Czernik 24 and Czernik 27 
  using the BOAO 1.8 m telescope and $BV$ CCD imaging data.
Since there have been no previous photometric study for these two clusters,
  we have determined, for the first time, the reddening, distance, metallicity, and
  age for these clusters and summarized them in Table 3.
Czernik 24 is found to be an old open cluster with age $1.8 \pm 0.2$ Gyr,
  metallicity [Fe/H]$=-0.41 \pm 0.15$ dex,
  distance modulus ${\rm (m-M)}_0 = 13.1 \pm 0.3$ mag (d$=4.1 \pm 0.5$ kpc), and
  reddening $E(B-V) = 0.54 \pm 0.12$ mag.
The parameters for Czernik 27 are estimated to be age = $0.63 \pm 0.07$ Gyr,
  [Fe/H]$=-0.02 \pm 0.10$ dex,
  ${\rm (m-M)}_0 = 13.8 \pm 0.2$ mag (d$=5.8 \pm 0.5$ kpc), and
  $E(B-V) = 0.15 \pm 0.05$ mag.
The metallicity and distance of the old open cluster Czernik 24
  obtained in this study are consistent with the
  general trend of the metallicity gradient along 
  the Galactocentric distance derived from 
  the photometric data of 51 old open clusters,
  which is found to be 
  $\Delta$[Fe/H]/$\Delta R_{gc}= -0.064 \pm 0.009$ dex kpc$^{-1}$.

% -------------------------------------------------- TABLE 3 START
\begin{table*}[t]
\begin{center}
{\bf Table 3.}~~Basic Information of Czernik 24 and Czernik 27 \\
\vskip 3mm
{\small
%{\footnotesize
\setlength{\tabcolsep}{1.2mm}
\begin{tabular}{lccl} \hline\hline
Parameter & Czernik 24 & Czernik 27 & Reference \\
\hline
Other names & C0552+208, OCL 472, Lund 200 & 
              C0700+064, OCL 526, Lund 284 & Lyng\r{a} 1987 \\ %Lynga
$\alpha_{2000}$ & 05$^h$ 55$^m$ 27$^s$             & 07$^h$ 03$^m$ 22$^s$ & This study \\
$\delta_{2000}$ & +20\arcdeg~ 52\arcmin~ 59\arcsec & +06\arcdeg~ 23\arcmin~ 47\arcsec & This study \\
$l$ & 188.\arcdeg06 & 208.\arcdeg58 & This study \\
$b$ & $-2.\arcdeg21$  & $+5.\arcdeg56$  & This study \\
Trumpler class      & III 1 m       & III 1 p & Lyng\r{a} 1987 \\ %Lynga
Reddening, $E(B-V)$               & $0.54 \pm 0.12$ mag  & $0.15 \pm 0.05$ mag & This study \\
Distance modulus, $V_0 - M_V$     & $13.1 \pm 0.3$ mag   & $13.8 \pm 0.2$ mag  & This study \\
Distance, d                       & $4.1 \pm 0.5$ kpc    & $5.8 \pm 0.5$ kpc   & This study \\
Galactocentric distance, $R_{gc}$ & $12.6 \pm 0.5$ kpc   & $13.9 \pm 0.5$ kpc & This study \\
Metallicity, [Fe/H]               & $-0.41 \pm 0.15$ dex & $-0.02 \pm 0.10$ dex & This study \\
Age, $t$                          & $1.8 \pm 0.2$ Gyr (${\rm log} ~t=9.25$) 
                                  & $0.63 \pm 0.07$ Gyr (${\rm log} ~t=8.80$) & This study \\
\hline
\end{tabular}
} %\small
\end{center}
\end{table*}
%-------------------------------------------------- TABLE 3 END

%acknowledgements   Acknowledgments
\vspace{4mm}
The authors thank the anonymous referee for improving this paper.
M.G.L. was supported in part by the ABRL (R14-2002-058-01000-0).

% =================

\end{document}